\documentclass[fleqn,twoside]{article}
\usepackage{espcrc2,amsmath}
\usepackage{graphicx}

\newcommand{\rf}[4]{{#1}{\bf #2}, (#4) #3}
\newcommand{\prd}{Phys.\ Rev.\ D }

\newcommand{\ppnp}{Prog.\ Part.\ Nucl.\ Phys.\ }

\newcommand{\epj}{Eur.\ Phys.\ J.\ }

\newcommand{\etal}{\emph{et al.}}
\newcommand{\punc}{, }
\newcommand{\last}{ and }
\newcommand{\rend}{.}

\newcommand{\psibar}{\overline{\psi}}

\newcommand{\mhat}{\widehat{m}}

\newcommand{\cond}{\langle\psibar\psi\rangle}

\newcommand{\eref}[1]{Eq.~(\ref{#1})}

\newcommand{\khat}{\hat{k}}
\newcommand{\ktilde}{\tilde{q}}

\newcommand{\Lqcd}{\ensuremath{\Lambda_{\text{QCD}}}}
\newcommand{\MSbar}{\overline{\text{MS}}}
\newcommand{\MOMtilde}{\widetilde{\text{MOM}}}
\newcommand{\oa}[1]{\ensuremath{{\cal O}(a^{#1})}}

\title{Infrared and ultraviolet properties of the Landau gauge quark
	propagator}
 
\author{Patrick O.~Bowman\address[ADL]{Department of Physics and CSSM,
University of Adelaide, Australia 5005}\thanks{Presented by
POB.},
Urs M.~Heller\address[PRD]{American Physical Society, One Research Road, 
Box 9000, Ridge, NY 11961-9000},
Derek B.~Leinweber\addressmark[ADL],
Anthony G.~Williams\addressmark[ADL] and
Jianbo Zhang\addressmark[ADL]}

\begin{document}

\begin{abstract}
We present a current summary of a program to study the quark propagator using 
lattice QCD.  We use the Overlap and ``Asqtad'' quark actions on a number
of lattice ensembles to assess systematic errors.  We comment on the place
of this work amongst studies of QCD Green's functions in other formulations.
A preliminary calculation of the running quark mass is presented.
\end{abstract}

\maketitle

\section{INTRODUCTION}
The quark propagator is a building block of Quantum Chromodynamics and has
been studied in numerous formulations~\cite{DSE,Dia02,Dor02}.  The lattice 
regularisation provides a method for a full nonperturbative calculation from 
first principles.  In the infrared, the quark propagator displays dynamical 
mass generation, but in the ultraviolet it takes its perturbative, running 
form.  

By computing the quark propagator on the lattice we can connect
with perturbation theory and with many models.  We can also contribute
to our understanding of the inner workings of the lattice simulations
themselves.  It has been seen, however, that detailed lattice studies of the
quark propagator require sophisticated 
actions~\cite{Sku01a,Sku01b,Bow02a,Bow02b,Bon02,overlap03}.

We present some results for the quark propagator in Landau gauge from a 
variety of lattice spacings, two different physical volumes and two different 
quark actions.
Specifically we use an improved Staggered action, ``Asqtad,''~\cite{Org99} 
and the Overlap quark action.  Results from Asqtad and Overlap actions are in 
good agreement.  We will compare our results to some model calculations and
compare our chiral extrapolation with recent Dyson--Schwinger equation 
predictions.   Not only do we see stable infrared results, but the 
ultraviolet data is good enough for us to attempt estimates of the chiral 
condensate and the running quark mass.  A much more detailed presentation
can be found in Ref.~\cite{springer}.  The quark propagator was also studied
in Laplacian gauge in Ref.~\cite{Bow02a}.

\section{THE LATTICE QUARK PROPAGATOR}

\begin{figure}[t]
\begin{flushleft}
\includegraphics[width=50mm,angle=90]{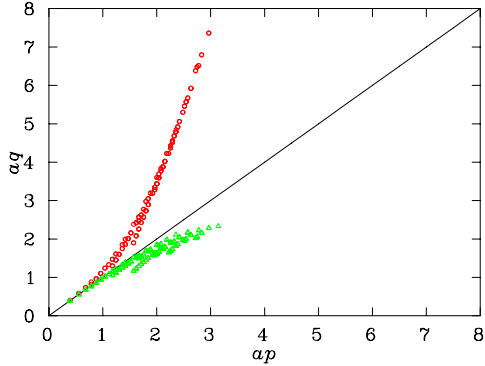}
\end{flushleft}
\vspace{-10mm}
\caption{Tree-level behaviour: $q$ vs.\ $p$ for the Overlap (top curve) and 
Asqtad (bottom curve) quark actions.  This is the inverse of the propagator 
for a massless quark.  The diagonal line is the continuum case.}
\label{fig:comp_tree}
\vspace{-3mm}
\end{figure}

In the (Euclidean) continuum, Lorentz invariance allows us to decompose the 
full quark propagator into Dirac vector and scalar pieces
\begin{equation}
S(p^2;m) = \frac{Z(p^2)}{i \gamma \cdot p + M(p^2)}.
\end{equation}
Asymptotic freedom means that at large momentum transfer, 
$p^2 \rightarrow \infty$, the quark propagator approaches its free form,
\begin{equation}
\label{eq:cont_free}
S(p^2;m) \rightarrow  \frac{1}{i\gamma \cdot p + m}, 
\end{equation}
where $m$ is a current quark mass.  On the lattice in the free case 
($U_\mu(x) = 1$), the quark propagator takes the form,
\begin{equation}
\label{eq:latt_free}
S(p_\mu;m) = \frac{1}{i \gamma \cdot q(p_\mu) + m},
\end{equation}
where the function $q_\mu(p_\mu)$ depends on the quark action.  For the 
Asqtad action it is~\cite{Bow02a},
\begin{equation}
\label{eq:mom_Asqtad}
q_\mu = \sin(p_\mu) \bigl[ 1 + \frac{1}{6} \sin^2(p_\mu) \bigr],
\end{equation}
and for the Overlap action it is~\cite{Bon02},
\begin{equation}
\label{eq:mom_Overlap}
q_\mu = 2 m_w \frac{ \ktilde_\mu \{ A + \sqrt{\ktilde^2 + A^2} \} }
	{ \ktilde^2 },
\end{equation}
where
\begin{gather}
A = -m_w + \frac{\khat^2}{2}  \nonumber\\
\ktilde_\mu = \sin(p_\mu)      \nonumber\\
\khat_\mu = 2 \sin(p_\mu / 2), \nonumber
\end{gather}
and $m_w$ is the regulator mass.  These functions are plotted in 
Fig.~\ref{fig:comp_tree}

The difference
between \eref{eq:cont_free} and \eref{eq:latt_free} is a pure lattice artefact.
For the Dirac vector piece this can be trivially corrected for by considering 
it to be a function not of $p_\mu$, but of $q_\mu$.  This only affects the 
ultraviolet behaviour of $Z$, and asymptotic freedom means that we may hope 
that this correction will persist in the interacting case.  This choice of
``kinematical momentum'' has been seen to be efficacious in the removal of
hypercubic artefacts~\cite{Bow02a} and the philosophy is the same as that
pursued with great success in studies of the gluon propagator~\cite{Bon01}.
Tree-level correction does not prescribe such a procedure for the Dirac scalar
piece, $M$, so we shall adopt an empirical approach.

\section{RESULTS}
\subsection{Lattice artefacts}

The simulation parameters are given in Table~\ref{tab:lattices}.  In all cases
an \oa{2} Symanzik improved gluon action was used.  The propagators were 
rotated to ${\mathcal O}(a^2)$-improved Landau gauge
\cite{Bonnet:1999mj} by standard methods. 

\begin{table}[tb]
\caption{Simulation parameters}
\label{tab:lattices}
\begin{tabular}{lllr}
\hline
 Action & $\beta$ & $a_\sigma$ (fm) & $L^3\times T$ \\
\hline
Asqtad  &   4.38  &   0.167  & $16^3 \times 32$ \\
Asqtad  &   4.60  &   0.124  & $12^3 \times 24$ \\
Asqtad  &   4.60  &   0.124  & $16^3 \times 32$ \\
Overlap &   4.286 &   0.190  &  $8^3 \times 16$ \\
Overlap &   4.60  &   0.124  & $12^3 \times 24$ \\
Overlap &   4.80  &   0.093  & $16^3 \times 32$ \\
\hline
\end{tabular}
\end{table}

\begin{figure}[tbh]
\begin{flushleft}
\includegraphics[width=50mm,angle=90]{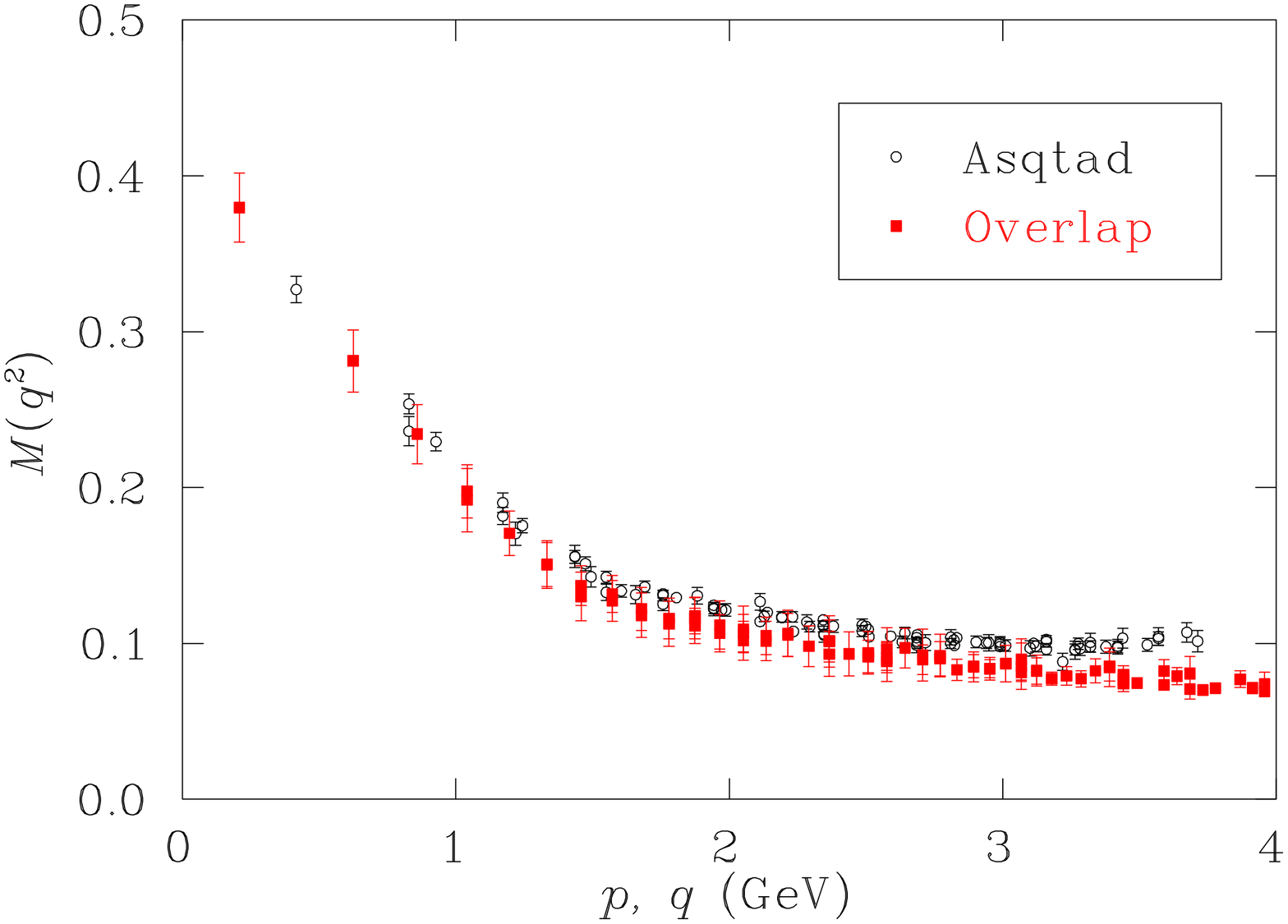}
\end{flushleft}
\begin{flushleft}
\includegraphics[width=50mm,angle=90]{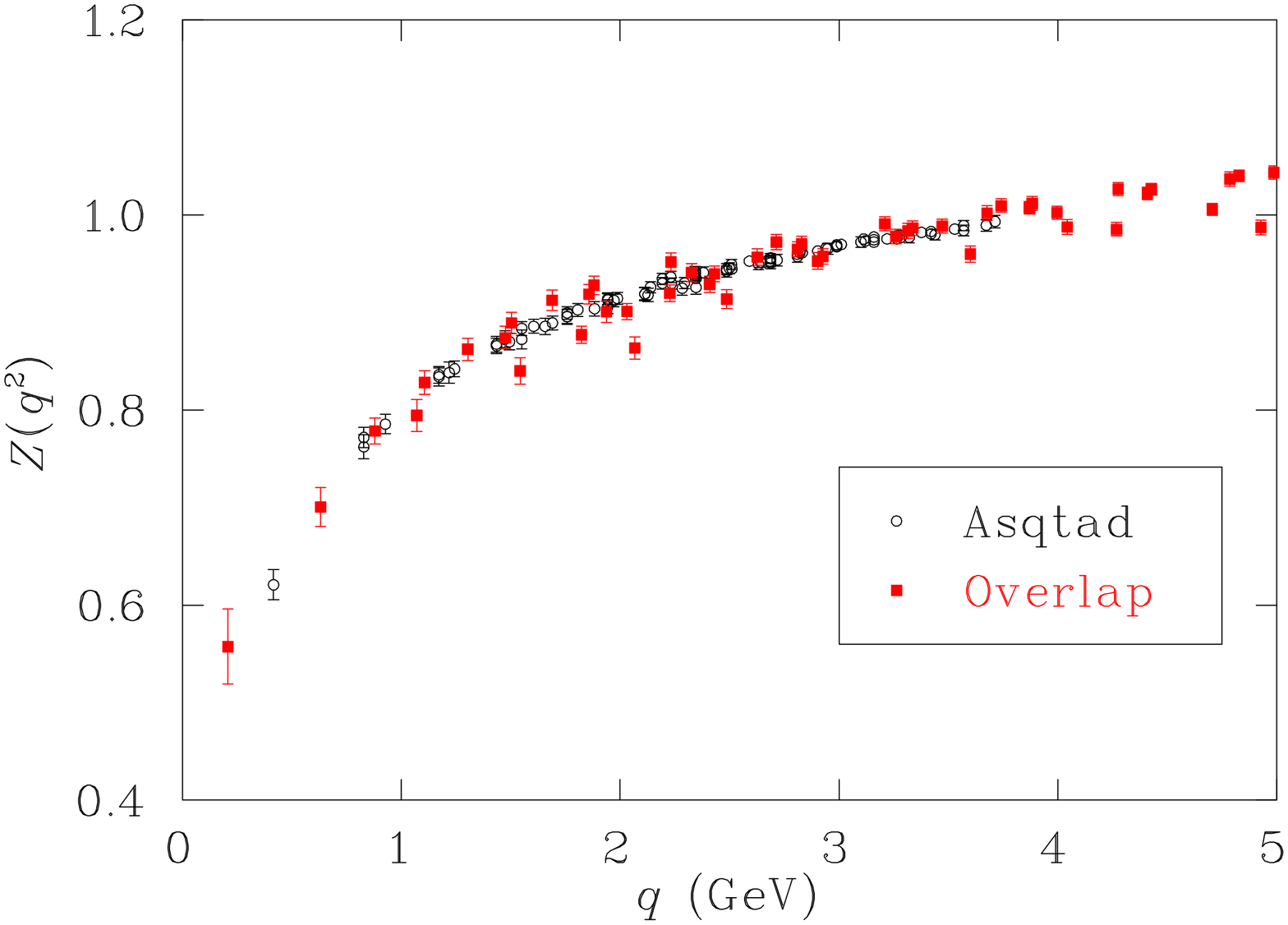}
\end{flushleft}
\vspace{-10mm}
\caption{Comparison of the actions: Quark mass function (top) and 
$Z$ function (bottom) for mass $m \simeq 80$ MeV at $\beta = 4.60$.  
These calculations were performed on a $12^3\times 24$ lattice.}
\label{fig:comp_actions}
\vspace{-3mm}
\end{figure}

As we have both Overlap and Asqtad results for one set of lattices we can
make a direct comparison.  This is shown in Fig.~\ref{fig:comp_actions} for
one choice of comparable quark masses.  $Z$ is shown as a function of $q$
-- chosen appropriately for each action -- and there is complete agreement,
within errors.  The Asqtad action has notably better rotational symmetry,
presumably reflecting the removal of \oa{2} errors.  Recently, one group has 
attempted to correct the poor rotational symmetry of the Overlap action by 
extrapolating in the hypercubic artefacts~\cite{Boucaud}.  The quark mass 
function for the Asqtad action is shown as a function of $q$, which has been 
seen to produce slightly better results~\cite{Bow02b}, but we use the standard
lattice momentum, $p$ for the Overlap action.  This choice produces the best 
agreement between the two actions.

\begin{figure}[tbh]
\begin{flushleft}
\includegraphics[width=50mm,angle=90]{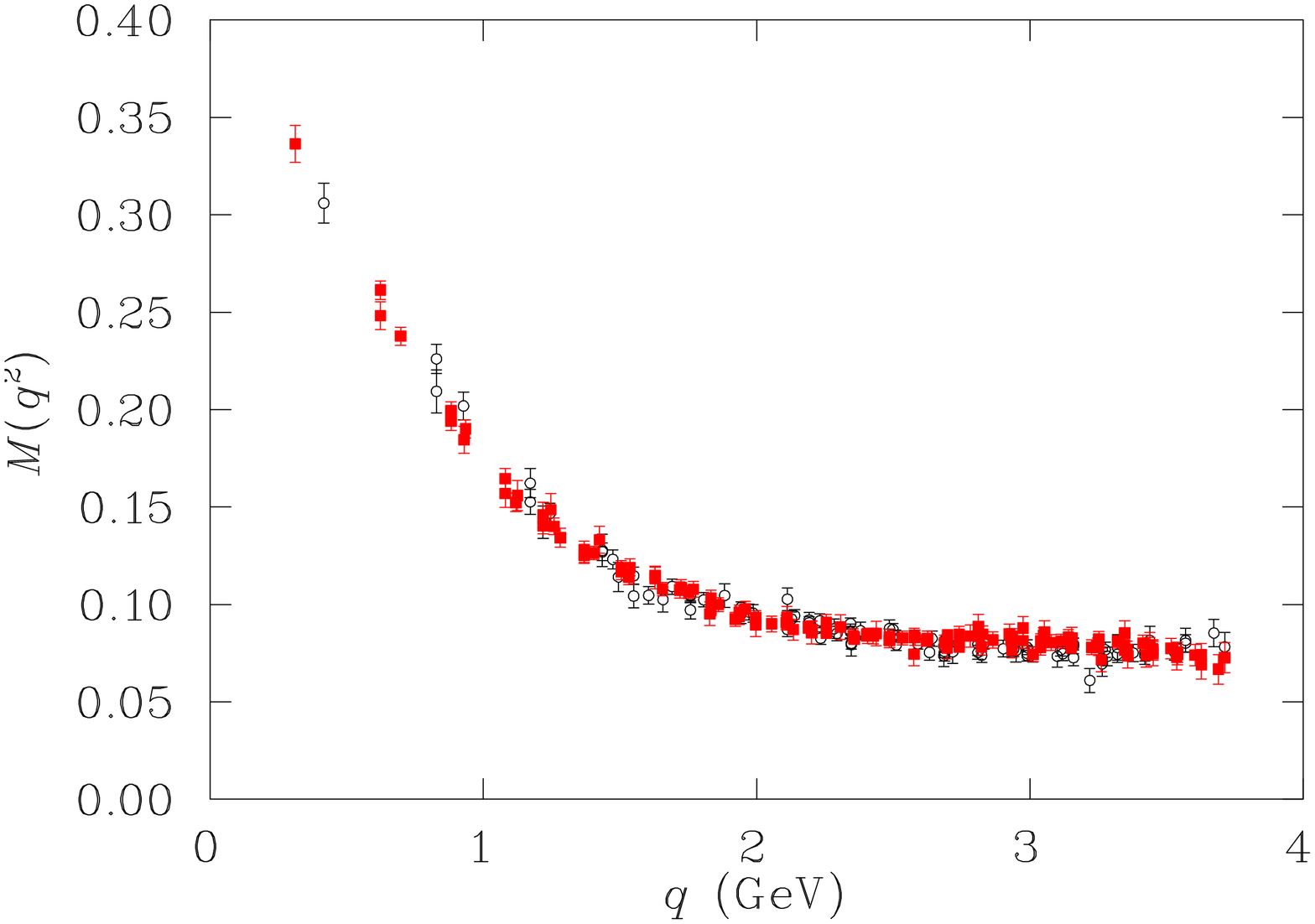}
\end{flushleft}
\begin{flushleft}
\includegraphics[width=50mm,angle=90]{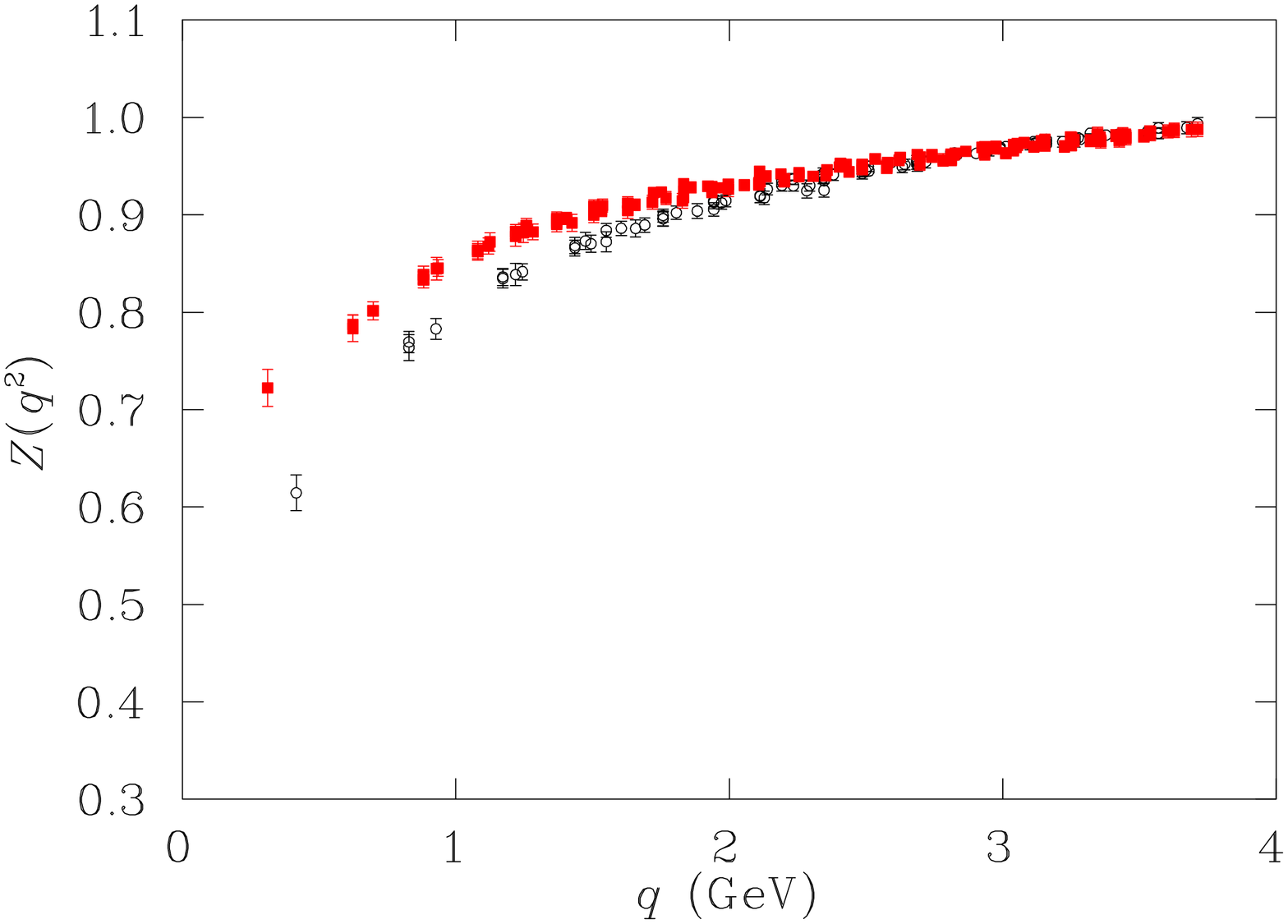}
\end{flushleft}
\vspace{-10mm}
\caption{Finite volume effects: Asqtad quark mass function (top) and 
$Z$ function (bottom) for mass $m \simeq 57$ MeV at $\beta = 4.60$.  
Comparison on $12^3\times 24$ lattice (open circles) and $16\times 32$ 
lattice (solid squares).}
\label{fig:finite_volume}
\vspace{-3mm}
\end{figure}

Figure~\ref{fig:finite_volume} shows the Asqtad quark propagator on two 
different size lattices with the same lattice spacing.  The smaller lattice has
a spatial length of about 1.5 fm, the larger about 2.0 fm.  The mass 
functions lie on top of each other, indicating that finite volume effects are 
small, if present.  For $Z$ however, there is a clear difference in the low 
momentum region.  Infrared suppression of $Z$ is a long-standing prediction of
Dyson--Schwinger equation studies and the larger lattice results are in good 
agreement
with recent DSE results~\cite{Bhagwat}, but at least some of the suppression
may be due to the finite volume.

\begin{figure}[tbh]
\begin{flushleft}
\includegraphics[width=50mm,angle=90]{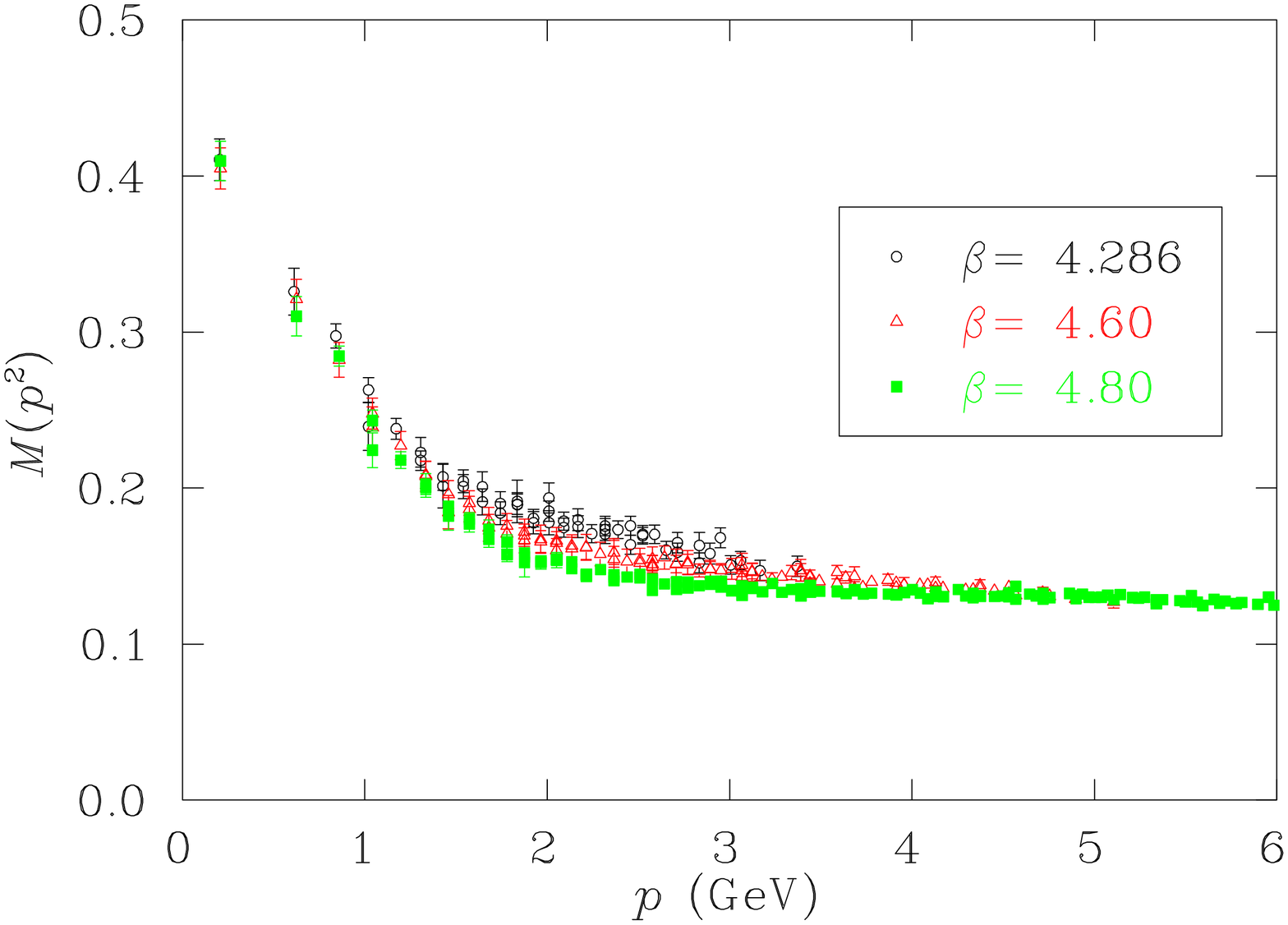}
\end{flushleft}
\begin{flushleft}
\includegraphics[width=50mm,angle=90]{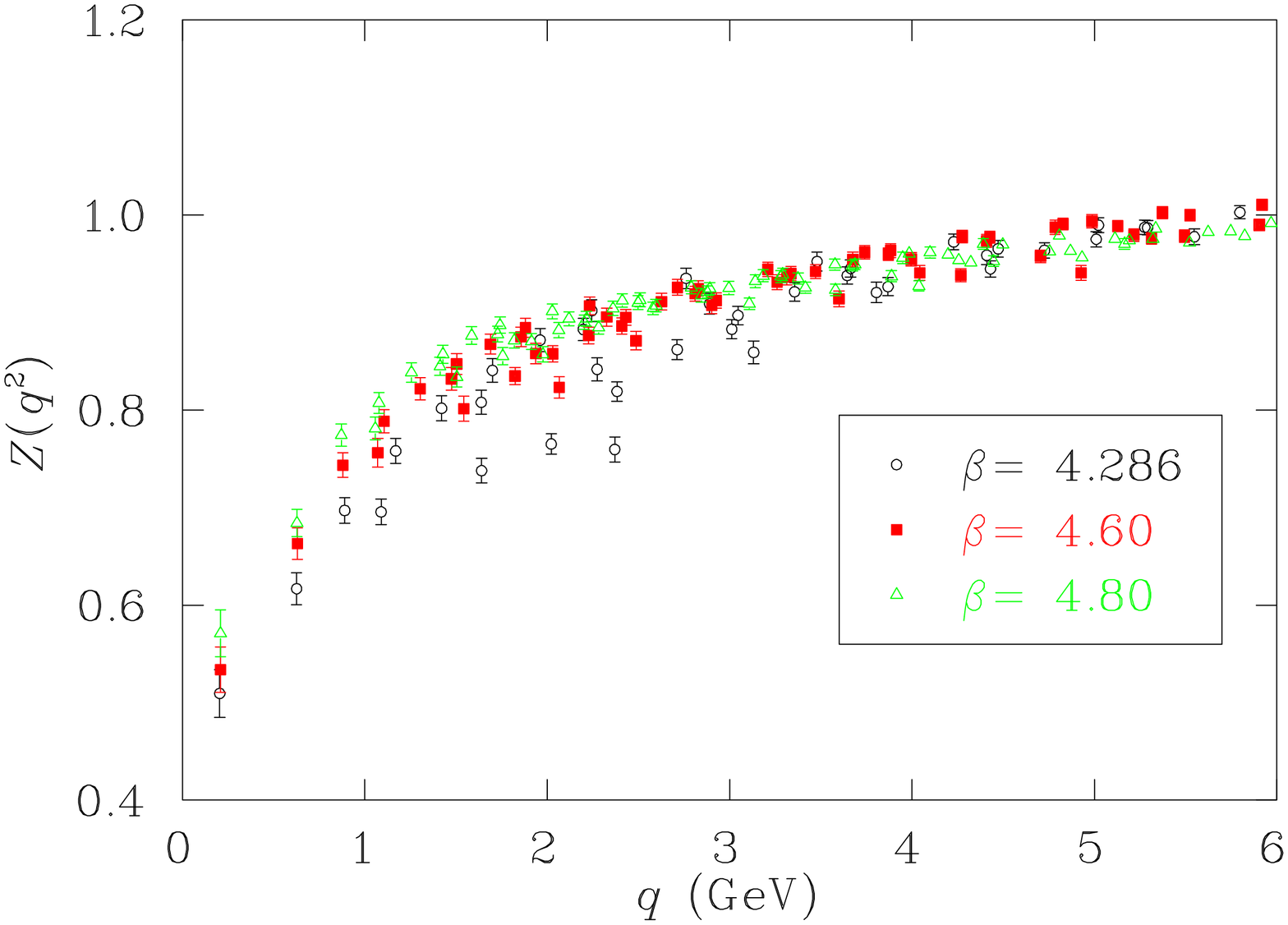}
\end{flushleft}
\vspace{-10mm}
\caption{Scaling: Quark mass (top) and $Z$ (bottom) for the Overlap action
at three lattice spacings.  Bare quark mass $m\simeq 106$ MeV.  $Z$ is shown
as a function of $q$ while $M$ is a function of $p$.  This is seen to provide
the best agreement between results on different lattice spacings.}
\label{fig:ovl_scaling}
\vspace{-3mm}
\end{figure}

As we have the Overlap quark propagator at three lattice spacings we can
examine its scaling behaviour.  This is shown in Fig.~\ref{fig:ovl_scaling}.
For both functions we see good consistency across these lattice spacings.
Once again, we have plotted $Z$ against $q$ and $M$ against $p$.  This gives
the best agreement between results at different lattice spacings.  The 
violation of rotational symmetry seen in $Z$ -- quite strong on the coarsest
lattice -- becomes smaller with the lattice spacing, but the overall form
is stable.  
The infrared behaviour of the mass function is almost unaffected by the change
in lattice spacing, but the ultraviolet shows some (not unexpected) 
sensitivity.

\begin{figure}[tbh]
\begin{flushleft}
\includegraphics[width=50mm,angle=90]{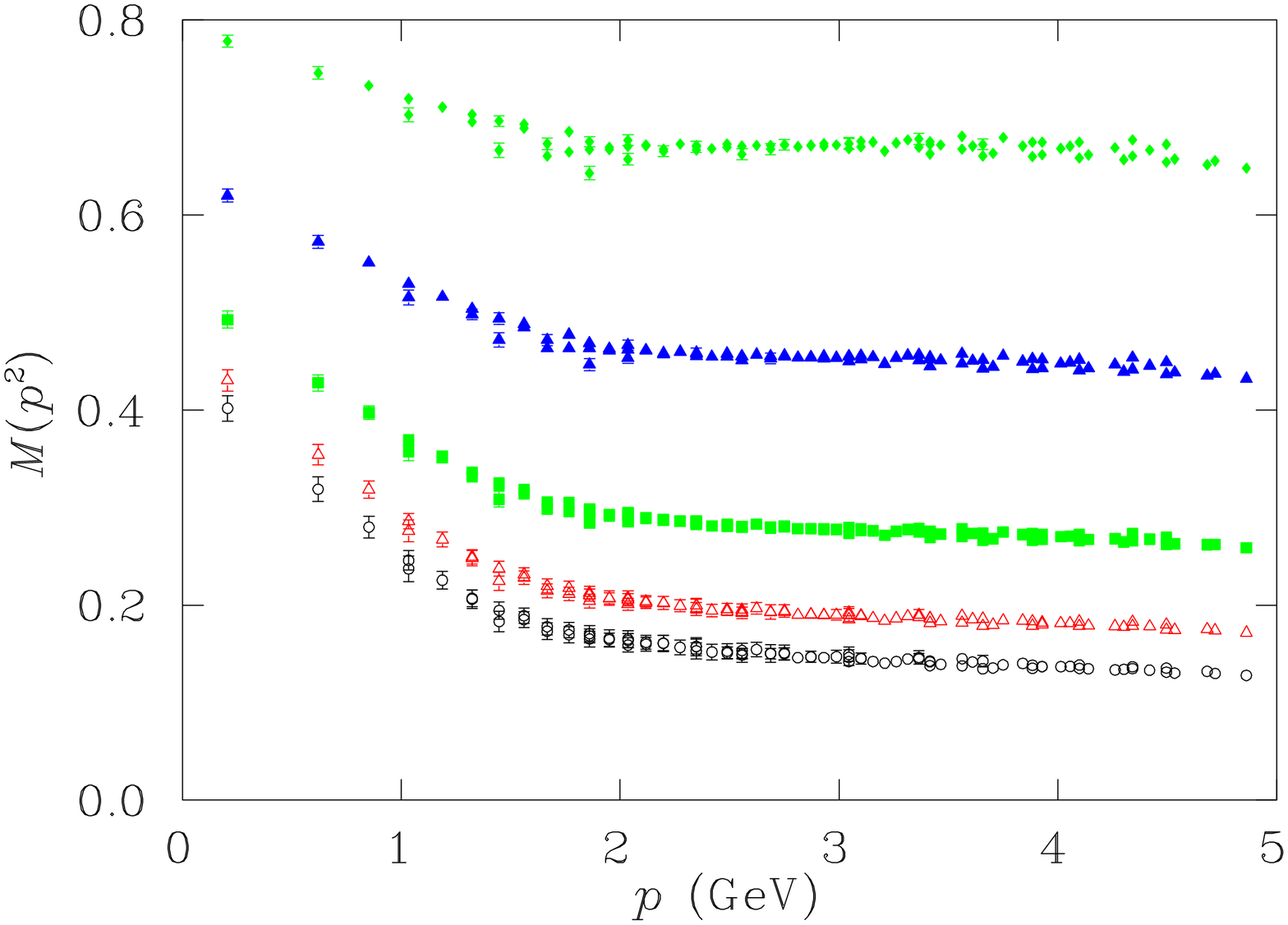}
\end{flushleft}
\begin{flushleft}
\includegraphics[width=50mm,angle=90]{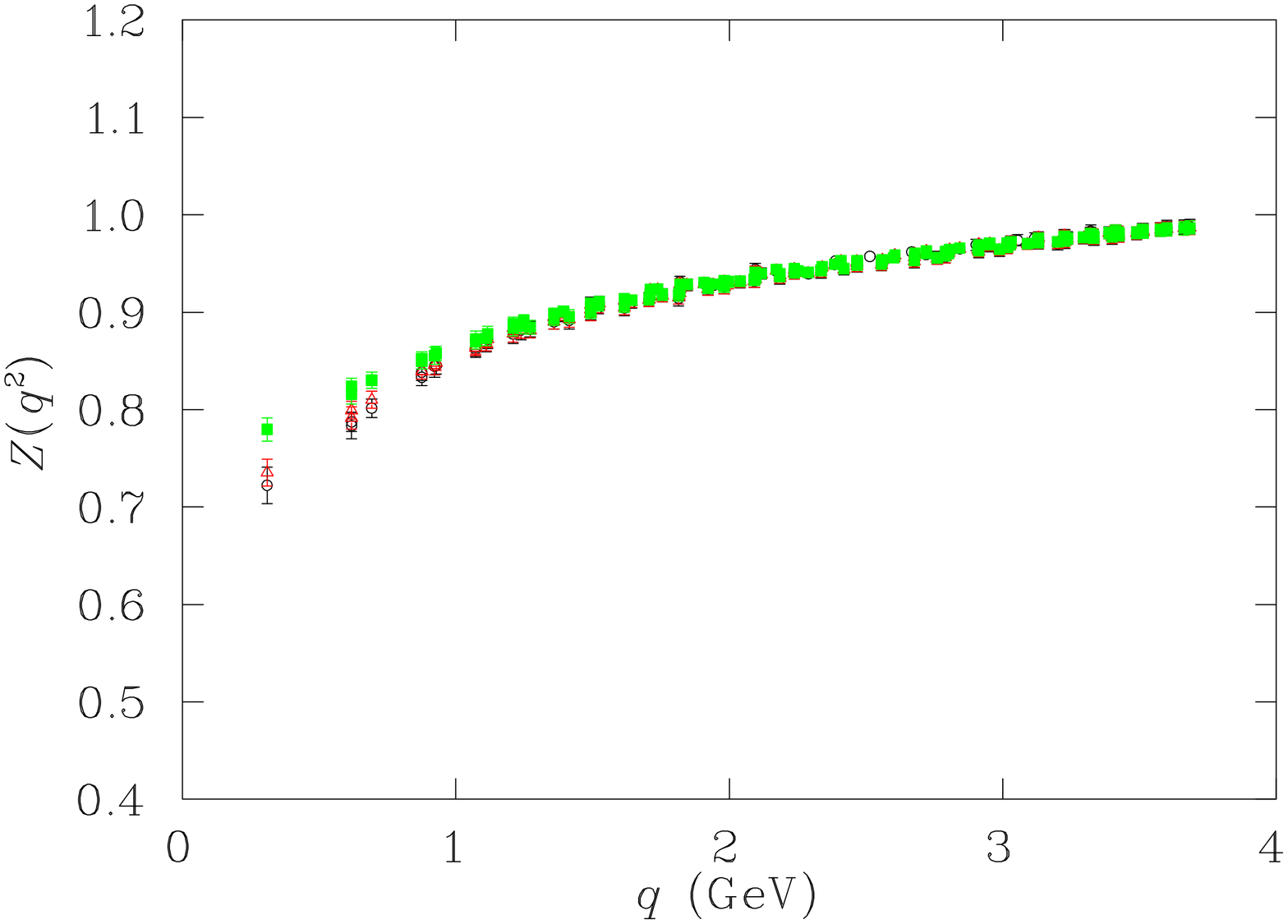}
\end{flushleft}
\vspace{-10mm}
\caption{Mass dependence: Asqtad Z function at $\beta = 4.60$, 
$16^3\times 32$, for bare quark masses $m \simeq 57$, 115 and 229 MeV.
There is some weak mass dependence in the infrared.
Overlap mass function at $\beta = 4.60$, $12^3\times 24$, for bare quark
masses $m \simeq 106$ - 531 MeV.  For large momentum, $M(p^2)$ is proportional
to the quark mass and as the mass increases, $M(p^2)$ becomes flatter.}
\label{fig:mass_dep}
\vspace{-3mm}
\end{figure}

The quark propagator is, of course, a function of the bare quark mass, and
this dependence is illustrated in Fig.~\ref{fig:mass_dep}.  For the $Z$ 
function we show results for three quark masses from the Asqtad action, and
indeed some mass dependence is observed in the infrared.  The heaviest 
quark has about four times the mass of the lightest one, so this dependence
is rather weak.  We expect the ultraviolet to be insensitive to the quark 
mass and this is in fact observed.  The ultraviolet tail of the mass 
function is proportional to the bare quark mass and this is observed in 
Fig.~\ref{fig:mass_dep} using the Overlap action.  As the mass 
increases the infrared enhancement becomes less significant.  It has already
been remarked elsewhere~\cite{Bow02b} that for small quark masses the 
infrared part of the mass function is insensitive to the quark mass, but at
around the strange quark mass the shape of the mass starts to change.  In the 
non-relativistic limit $M(p^2)$ would be a constant.

\subsection{Comparison with other methods}
 
\begin{figure}[tbh]
\begin{flushleft}
\includegraphics[width=50mm,angle=90]{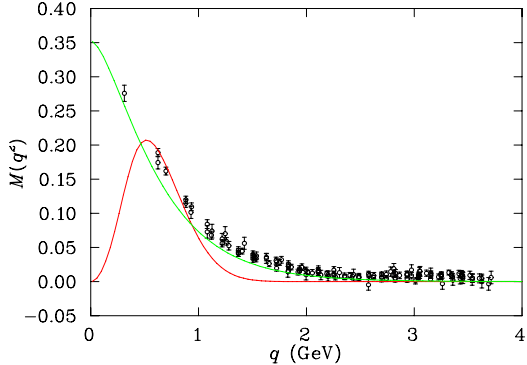}
\end{flushleft}
\vspace{-10mm}
\caption{Comparison with two models: The mass function that vanishes at
zero four--momentum is from a model of Dorokhov and Broniowski~\cite{Dor02},
the other is from an instanton calculation by Diakonov~\cite{Dia02}.}
\label{fig:comp_models}
\vspace{-3mm}
\end{figure}

We extrapolate the quark mass function to the chiral limit using a quadratic
fit.  One reason for this is that it is a convenient quantity to compare with
other calculations.
In Ref.~\cite{Dor02} the authors suggested a novel form of the quark mass 
function which actually vanishes at zero four-momentum.  The ansatz
\begin{equation}
M(q^2) = \frac{1 - \sqrt{1 - q^2 Q^2(q^2)}}{Q(q^2)},
\label{eq:CIM}
\end{equation}
where
\begin{equation}
Q(q^2) = \frac{-\cond}{2N_c}\frac{16\pi^2}{\lambda^4} e^{-q^2 / \lambda^2},
\end{equation}
based on the Constrained Instanton Model and QCD sum rules with non-local
condensates, is one of the curves shown in Fig.~\ref{fig:comp_models}.  This 
form is not favoured by the lattice data.  It is 
fascinating that such a form should be consistent with the phenomenological 
constraints discussed in Ref.~\cite{Dor02} and demonstrates the utility of 
first-principles calculations.

Figure~\ref{fig:comp_models} also shows a regular instanton model calculation
from Ref.~\cite{Dia02}. There the quark mass function takes the 
form
\begin{align}
M(q) & = M(0) F^2(z) \\
F(z) & = 2z \Bigl( I_0(z)K_1(z) - I_1(z)K_0(z) \nonumber\\
     & \qquad - I_1(z)\frac{K_1(z)}{z} \Bigr) \nonumber\\ 
z    & = q \frac{\rho}{2} \nonumber
\end{align}
where the I's and K's are the modified Bessel functions.  The parameters
$M(0) = 350$ MeV and $\rho^{-1} = 600$ MeV are derived from the standard
values of the instanton\index{instanton} ensemble.  This two parameter form 
has a good correspondance with the lattice data.

The interaction between lattice and DSE studies has been particularly active
recently. See Refs.~\cite{Bhagwat,Oet03,Iida,Alkofer} for a few examples.
Amongst other things, this is providing valuable insight about the 
validity of various DSE truncation schemes.  In return, DSE calculations
can be done at quark masses and momenta difficult to access on the lattice 
and can also provide information on the time-like region.

In Ref.~\cite{Bhagwat} there is a comparison of the lattice quark mass function
at various quark masses with a form derived from the Dyson--Schwinger equation.
The DSE describes the lattice data well at finite quark mass, but there is
some difference in the chiral limit.  In particular the DSE solution changes
more rapidly with momentum around one GeV.  This suggests some non-linear
behaviour at intermediate momenta and small quark masses, not captured by
out simple extrapolation.  While such behaviour is certainly not surprising, 
it is also not well understood.

\section{ASYMPTOTIC BEHAVIOUR}

As we have already remarked, although the quark propagator is gauge dependent,
and hence not observable, certain gauge invariant quantities can be extracted
from it directly.
In the continuum, the quark mass function has the asymptotic form~\cite{DSE},
\begin{multline}
\label{eq:mass_asymp}
M(q^2) \underset{q^2\rightarrow \infty}{=} 
       \frac{c}{q^2} \bigl[ \ln(q^2 / \Lqcd^2) \bigr]^{d_M-1} \\
         + \frac{\mhat}{\bigl[ \ln(q^2/\Lqcd^2) \bigr]^{d_M}}
\end{multline}
where $\mhat$ is the RGI (renormalisation 
group invariant) mass and the anomalous dimension of the quark mass is
$d_M = 12/(33 - 2N_f)$ for $N_f$ quark flavours (zero here).  In this form 
the fact that the mass function is independent of the choice of 
renormalisation point is manifest.  The first term in \eref{eq:mass_asymp} is 
of nonperturbative origin and, in the chiral limit, a non-zero value for $c$ 
indicates the spontaneous breaking of chiral symmetry.  In this case it is 
related to the usual chiral condensate by
\begin{equation}
\label{eq:cond_asymp}
c \simeq - \frac{4\pi^2 d_M}{3} 
         \frac{\langle\psibar \psi\rangle}{[\ln(\mu^2 / \Lqcd^2)]^{d_M}},
\end{equation} 
where $\mu^2$ is a choice of renormalisation point.  
\index{chiral condensate}%
The second term in
\eref{eq:mass_asymp} comes from perturbation theory (one-loop) and breaks
chiral symmetry explicitly.  The RGI quark mass in 
\eref{eq:mass_asymp} can be replaced by the running quark mass,
\index{running mass}%
\begin{equation}
\label{eq:running}
\mhat = m(\mu^2)[ \ln(\mu^2/\Lqcd^2) ]^{d_M}.
\end{equation}
Table \ref{asympt} summarizes results for a fit of
Eq.~(\ref{eq:mass_asymp}) to our lattice results.

\begin{table}[tb]
\caption{\label{tab:asymp_fits}Results for the fit to \eref{eq:mass_asymp}.  
The fit region corresponds to 1.9-2.9 GeV, which includes 51 data
points, and was chosen to minimise the $\chi^2/$dof.  In the chiral limit this gives a
value for the condensate of $(-\cond)^{1/3} = 274(24)$ MeV at the 
renomalisation point $\mu = 2$ GeV.}
\label{asympt}
\begin{tabular}{lllc}
\hline
$am$   &   $a^3c$   & $\mhat$ (MeV) & $\chi^2$ / d.o.f. \\
\hline
    0  &  0.018(6)  &      0        & 0.52 \\
0.012  &  0.019(3)  &     28(2)     & 0.57 \\
0.018  &  0.014(12) &     44(5)     & 0.54 \\
0.024  &  0.0096(47)&     59(3)     & 0.52 \\
0.036  &  0.0026(104)&    90(5)     & 0.49 \\
0.048  &  0.086(3)  &    137(4)     & 0.46 \\
0.072  &  0.130(3)  &    204(5)     & 0.40 \\
0.108  &  0.190(3)  &    300(5)     & 0.31 \\
0.144  &  0.251(5)  &    397(8)     & 0.25 \\
\hline
\end{tabular}
\end{table}

Having fit the asymptotic form to the quark mass function we should also be
able to calculate the quark mass, but there is one piece of information
missing from Table~\ref{tab:asymp_fits}.  The effect of the bare quark mass 
on physical quantities is theory dependent; we have no \emph{a priori}
knowledge of where the physical point lies.  We can match a physical quantity
to its real world value and thus extract the relevant bare mass for that 
particular action, with those particular lattice parameters.  In this case
we choose the Asqtad quark action on the $16^3\times 32$ lattice at 
$\beta=4.60$ and set the ``physical'' bare quark mass by extrapolation to the
physical pion mass.

We fit the five lowest $\pi$ masses to the form
\begin{equation}
\label{eq:chiral_pi}
am_\pi = \sqrt{a^2Bm},
\end{equation}
the lowest order result from chiral perturbation theory.  For the higher 
masses, the data deviates from this form due to the contribution from higher
order terms; in the heavy quark limit $m_\pi \propto m$.  The U(1) chiral
symmetry of staggered quarks ensures that there is no additive mass 
renormalisation.  To find the physical point we find the bare mass 
corresponding to $m_\pi = 140$ MeV for each jackknife bin and get 
$am = 0.00215(1)$.  Then we fit the lowest four RGI masses in 
Table~\ref{tab:asymp_fits} to the form
\begin{equation}
a\mhat = Aam
\end{equation}
and find $A = 1.57(16)$.
Putting these together we get
\begin{equation*}
\mhat = 5.3(6) \text{MeV}.
\end{equation*}
The RGI quark mass in \eref{eq:mass_asymp} is related to 
the running quark mass through \eref{eq:running}.
\index{running mass}%
Thus by choosing the renormalisation point $\mu = 2$ GeV and inserting 
$\Lqcd^{\MSbar} = 239$ MeV, we get
\begin{equation*}
m(\mu=2\text{ GeV})^{\MSbar} = 3.1(4) \text{MeV},
\end{equation*}
or with $\Lqcd^{\MOMtilde} = 691$ MeV,
\begin{equation*}
m(\mu=2\text{ GeV})^{\MOMtilde} = 4.0(5) \text{MeV}.
\end{equation*}
This result compares favourably with the Particle Data Group's number for the 
average of the up and down quark masses~\cite{PDG02},
\begin{equation*}
\overline{m}(\mu=2\text{ GeV})^{\MSbar} = 2.5-5.5 \text{MeV}.
\end{equation*}

The curious behaviour of the condensate term at small masses is easily 
explained.  The asymptotic behaviour of the quark mass function quickly 
becomes dominated by the explicit chiral symmetry breaking mass; at the same
time, large fluctuations make the mass function relatively noisy at these 
masses.  Together, that means that sensitivity of the fit to the first term in 
\eref{eq:mass_asymp} is lost.  At larger masses, the signal becomes cleaner 
and statistical errors shrink, restoring stability to the fits.  One way of
countering this effect would be to fix $ac$ for each light mass to some 
reasonable value given the chiral and heavy quark values.  The
difference to the end result would be small, but a future study should consider
such systematics.  

This calculation is, of course, quenched.  There are also systematic errors
associated with the two extrapolations.  Using other observables to set the 
scale -- such as the $\rho$ and nucleon masses, possibly supported by the 
appropriate finite-range regulated
\cite{Young:2002ib,Leinweber:2003dg} staggered chiral 
perturbation theory \cite{Aubin:2003mg,Aubin:2003uc}
-- would help us 
understand the systematic uncertainty associated with the extrapolation.  
Such an approach could also provide a handle on the uncertainty associated 
with quenching.  Finally, we should realise that this is a rather audacious 
approach: we are exploiting ultraviolet physics on a relatively coarse 
lattice.  It is very encouraging, however, that we are able to get such a
promising result with such modest resources.  The choices of improved quark and
gauge actions are essential to this.

\section{CONCLUSIONS}

A comprehensive understanding of the quenched quark propagator in
Landau gauge is now in place.  Overlap and Improved Staggered actions
produce propagators that are well behaved over a wide range of
momenta.  Despite this, more points in the deep infrared are required
to pin down precisely the behaviour near zero four-momentum.  Also,
the role of finite volume effects in the $Z$ function at small quark
masses needs further investigation.  Calculations using dynamical
configurations are underway.

\section*{ACKNOWLEDGMENTS}

The authors would like to thank the organisers of the Lattice Hadron Physics
 workshop (Cairns, 2003) for a great workshop.


\end{document}